\documentclass[twocolumn,showpacs,preprintnumbers,amsmath,amssymb,floatfix]{revtex4}

\usepackage{graphicx}
\usepackage{dcolumn}
\usepackage{bm}

\newcommand{\beq}{\begin{equation}}

\newcommand{\eeq}{\end{equation}}
\newcommand{\bey}{\begin{eqnarray}}
\newcommand{\eey}{\end{eqnarray}}

\begin{document}

\title{Quark matter as dark matter in modeling galactic halo}

\author{Farook Rahaman}
\email{rahaman@iucaa.ernet.in} \affiliation{Department of
Mathematics, Jadavpur University, Kolkata 700032, West Bengal,
India}
\author{P.K.F. Kuhfittig}
\email{kuhfitti@msoe.edu} \affiliation{Department of Mathematics,
Milwaukee School of Engineering, Milwaukee, Wisconsin 53202-3109,
USA}
\author{Ruhul Amin}
\email{ramin@ewubd.edu} \affiliation{Department of Electronics
and Communications Engineering, East West University, Dhaka 1212,
Bangladesh}
\author{Gurudas Mandal}
\email{gdmandal@ewubd.edu} \affiliation{Department of Electronics
and Communications Engineering, East West University, Dhaka 1212,
Bangladesh}
\author{Saibal Ray}
\email{saibal@iucaa.ernet.in}\affiliation{Department of Physics,
Government College of Engineering and Ceramic Thechnology, Kolkata
700010, West Bengal, India}
\author{Nasarul Islam}
\email{nasaiitk@gmail.com}\affiliation{Department of Mathematics,
Danga High Madrasah, Kolkata 700103, West Bengal, India}

\date{\today}

\begin{abstract}
Considering the flat rotation curves as input and treating the
matter content in the galactic halo region as quark matter, we
have found out a background spacetime metric for the region of the
galactic halo. We obtain fairly general conditions that ensure
that gravity in the halo region is attractive. We also investigate
the stability of circular orbits, along with a different role for
quark matter. Bag-model quark matter meeting these conditions
therefore provides a suitable model for dark matter.
\end{abstract}


\maketitle

\section{Introduction}
The flatness of the galactic rotation curves of neutral hydrogen
clouds in the outer regions of galaxies has led to the hypothesis
that galaxies and even clusters of galaxies are pervaded by dark
matter \cite{jO32,fZ33,fZ37}. To explain the observed constant
velocity, it is assumed that the decrease in the energy density
is proportional to $1/r^2$, where $r$ is the distance from the
center of the galaxy.

In special connection to these galactic halo and dark matter we
would like to mention some of the previous works in the sequel as
follows. In an investigation under the framework of brane-world
models it has been shown by Rahaman et al. \cite{Rahaman2008} that
the observed rotation curves result solely from non-local effects
of gravitation, such as dark radiation and dark pressure, and does
not invoke any exotic matter field. Nandi et al. \cite{Nandi2009a}
have considered several aspects of the 4d imprint of the 5d bulk
Weyl radiation by combined measurements of rotation curve and
lensing effect. In another case the energy density attached to
noncommutative geometry has been assumed to diffuse throughout a
region which is responsible for producing stable circular orbits
and attractive gravity \cite{Rahaman2012}. For some other notatble
works in this line see Refs.
\cite{Matos2000,Nandi2009b,Rahaman2010}.

A number of candidates for this dark matter has been put forward
\cite{ESM90,aP04,Khlopov2011}. In the present paper we propose
that quark matter \cite{Heinz2000,Schaefer2003,Muller2005} is such
a candidate, as previously suggested by several workers
\cite{PSS,Forbes2010,Weissenborn2011}. The fundamental particles
of quark matter do not ordinarily exist as free particles since
they are bound together by the strong interaction. However, quark
matter is believed to exist at the center of {\it neutron stars}
\cite{PSS}, in {\it strange stars}
\cite{Drake2002,Jaikumar2006,Rodrigues2011,Bordbar2011},
or even as small pieces of {\it strange matter}
\cite{Madsen1999,Weber2005}.

Regarding the origin and survival of quark matter, it is believed
that the Universe underwent a quark gluon phase transition a few
microseconds after the big bang
\cite{Bhattacharyya1999,Bhattacharyya2000,Bhattacharyya2003}. In
fact, in 1984, Witten \cite{Witten1984} proposed that such a
transition at a critical temperature $T_C\equiv 100-200$ MeV could
have led to the formation of quark nuggets made of u, d, and s
quarks at a density that would be larger than normal nuclear
matter density. Laboratory experiments (such as CERN LHC) with
relativistic nuclei aim to recreate the conditions similar to
those encountered before and in the early hadronisation period
\cite{EMU1991}. As the expanding Universe cools, the hot
quark-gluon-plasma (QGP) freezes slowly into individual hadrons
\cite{Fromerth2005}.

The survival of such nuggets has been the subject of many
investigations
\cite{Madsen1986,Madsen1991,Sumiyoshi1991,Bhattacharjee1993}.
According to Bhattacharyya et al. \cite{Bhattacharyya1999}, not
only does a large number of stable quark nuggets exist in the
present Universe, but quark matter could be a viable candidate for
cosmological dark matter, as already noted. Quantum Chromo
Dynamics (QCD), the most accepted theory of strong interaction
predicts that under extreme condition a hadronic system can
undergo a phase transition from color confined hadronic matter to
the QGP phase. In Astrophysics, it is speculated that the core of
neutron stars may consist of cold QGP \cite{Walker1991}.

Since we are taking the flat rotation curves as input, the
circular orbit is assumed accordingly. The problem is thereby
reduced to usages of the MIT bag model \cite{Chodos1974} to
determine conditions ensuring that gravity in the halo region is
attractive. Under those conditions, quark matter may be considered
as one of the suitable candidates for dark matter. This is the
motivation behind the present work.

The investigations are organized as follows: in Sec. 2 we provide
basic equations whereas in Sec. III we have considered the
galactic rotation curves as an input. The solutions of the
equations have been found out for the cases (i)
 non-interacting two-fluid model, and (ii) interacting two-fluid model in Section 4.
In Secs. 5, 6 and 7 we have addressed the question of an
attractive gravity for structure formation in the galactic region,
stability of circular orbits and dual role of quark matter,
respectively. Sec. 8 is devoted for some concluding remarks on the
model.

\section{Basic Equations}
In this paper the metric for a spherically symmetric spacetime
is taken to be
\begin{equation}
ds^2=-e^{\nu(r)}dt^2 + e^{\lambda(r)}dr^2+r^2
  d\Omega^2, \label{line1}
\end{equation}
where $d\Omega^2=d\theta^2+{sin}^2\theta\, d\phi^2$. We are using
here geometrized units in which $G=c=1$.

The energy-momentum tensor of the two-fluid model is given by
\begin{eqnarray}
T^{0}_{0} \equiv \rho_{eff} &=& \rho + \rho_{q}, \label{eq2}\\
T^{1}_{1} = T^{2}_{2}  \equiv - p_{eff} &=& -(p +p_{q}),
\label{eq3}
\end{eqnarray}
In the above two equations (\ref{eq2}) and (\ref{eq3}) $\rho$ and $p$
correspond to the respective energy
density and pressure of the baryonic matter, whereas
$\rho_{q}$ and $p_q$
to the respective {\it dark} energy density and pressure due to quark
matter. The left-hand sides of equations (\ref{eq2}) and (\ref{eq3})
are the effective energy density and pressure, respectively,
of the composition.

The Einstein field equations are listed next:
\begin{eqnarray}
\label{eq4} 8\pi\left(\rho+\rho_{q}\right) &=&
e^{-\lambda}\left(\frac{\lambda^\prime}{r}-\frac{1}{r^2}\right) +
\frac{1}{r^2},\\
\label{eq5} 8\pi\left(p+p_{q}\right) &=&
e^{-\lambda}\left(\frac{\nu^\prime}{r}+\frac{1}{r^2}\right) -
\frac{1}{r^2},\\
\label{eq6} 8\pi\left(p+p_{q}\right) &=&
\frac{e^{-\lambda}}{2}\left[\frac{{(\nu^\prime})^2 -
\lambda^{\prime}\nu^{\prime}}{2}
+\frac{\nu^\prime-\lambda^\prime}{r}+\nu^{\prime\prime}\right],
\end{eqnarray}
since $T^{1}_{1}=T^{2}_{2}$.

In the MIT bag model, the quark matter equation of state (EoS-1) has the
simple linear form
 \begin{equation}p_q= \frac{1}{3}(\rho_q-4B), \label{eq7}
\end{equation}
where $B$, the bag constant, is in units of $MeV/(fm)^3$
\cite{Chodos1974,Rahaman2011}.

For normal matter, we use the following equation of state (EoS-2):
\begin{equation}
  p = m \rho, \label{eq8}
\end{equation}
where $0<m <1$.

Since we are assuming the pressure to be isotropic, therefore, the
conservation equation is

\begin{equation}
 \frac{d(p_{eff})}{dr}   +\frac{1}{2}
\nu^\prime\left(\rho_{eff} +p_{eff}\right) = 0  \label{eq9}.
\end{equation}

\section{Galactic rotation curves}
The observed flat rotation curves are often considered as evidence for
the existence of dark matter. In such galaxies the neutral hydrogen
clouds, observed at large distances from the center, are treated
as test particles moving in circular orbits due to the gravitational
effects of the halo. To derive the tangential velocity of such
circular orbits, we start with the line element (\ref{line1}) and
then observe that the Lagrangian for a test particle takes the
following form:
\begin{equation}\label{E:Lagrangian2}
2\mathcal{L}=-e^{\nu(r)}\dot{t}^{2}+e^{\lambda(r)}\dot{r}^{2}%
+r^{2}\dot{\Omega}^{2},
\end{equation}
where as usual
$\dot{\Omega}^2=\dot{\theta}^2+{sin}^2\theta\,\dot{\phi}^2$. The
overdot denotes differentiation with respect to affine parameter
$s$.

Since the metric tensor coefficients do not depend explicitly on
$t$, $\theta$, $\phi$, or $\Omega$, the Euler-Lagrange equation
yields directly the conserved quantities $E$ and $L$, the energy
and total momentum, respectively: $E=-e^{\nu(r)}\dot{t}$,
$L_{\theta}=r^2\dot{\theta}$, and
$L_{\phi}=r^{2}{sin}^{2}\theta\,\dot{\phi}$. So the square of the
total angular momentum is $L^{2}=
{L_{\theta}}^{2}+(L_{\phi}/\sin\theta) ^{2}$. It is actually more
convenient to use Eq. (\ref{E:Lagrangian2}), as considered by
B{\"o}hmer, Harko and Lobo \cite{BHL08}. Then the total angular
momentum becomes $L=r^2\dot{\Omega}$.

With the use of the conserved quantities $E$ and $L$ and
the norm of the four-velocity $u^{\mu}u_{\mu}=-1$, the
geodesic equation becomes
\begin{equation}\label{E:geodesic1}
  -1=-e^{\nu(r)}\dot{t}^2+e^{\lambda(r)}\dot{r}^2
  +r^2(\dot{\theta}^2+{sin}^2\theta\,\dot{\phi}^2).
  \end{equation}
Hence we get
\begin{equation}\label{E:V1}
  e^{\nu(r)+\lambda(r)}\dot{r}^2+e^{\nu(r)}
  \left(1+\frac{L^2}{r^2}\right)=E^2.
\end{equation}
The equation of motion
\begin{equation}
  \dot{r}^{2}+V(r)=0, \label{E:motion}
\end{equation}
now yields the potential as
\begin{equation}\label{E:V2}
V(r)=-e^{-\lambda(r)}\left(
e^{-\nu(r)}E^{2}-\frac{L^{2}}{r^{2}}-1\right).
\end{equation}
Following the work of Nucamendi, Salgado and Sudarsky \cite{NSS01},
as it is more convenient to use the effective potential
$V_{eff}(r)$, we write Eq. (\ref{E:V1}) in the form
\begin{equation}\label{E:V3}
   e^{\lambda(r)}\dot{r}^2+1+\frac{L^2}{r^2}
    -e^{-\nu(r)}E^2=0.
\end{equation}

Recall that for the case of circular stable orbits, the
effective potential must satisfy the following conditions:
(1) $\dot{r}^2=0$, (2) $\frac{\partial V_{eff}}{\partial r}=0$,
and (3) $\frac{\partial^2 V_{eff}}{\partial r^2}>0$.   By
using
\begin{equation}\label{E:Veff}
  V_{eff}(r)=1+\frac{L^2}{r^2}-e^{-\nu}E^2,
\end{equation}
the first condition gives directly
\begin{equation}\label{E1}
  E^2=e^{\nu}\left(1+\frac{L^2}{r^2}\right)
\end{equation}
from Eq. ({\ref{E:V3}), while the second condition yields
\begin{equation}\label{E:L1}
  \frac{L^2}{r^2}=\frac{1}{2}\,r\,\nu'e^{-\nu}E^2.
\end{equation}
These equations can also be rewritten as
\begin{equation}\label{E:E2}
  E^2=\frac{e^{\nu}}{1-\frac{1}{2}\,r\,\nu'}
\end{equation}
and
\begin{equation}\label{E:L2}
  L^2=\frac{\frac{1}{2}\,r^3\nu'}{1-\frac{1}{2}\,r\,\nu'}.
\end{equation}

Before considering the third condition, $V_{eff}(r)_{rr}>0$,
we need to obtain an expression for the tangential velocity
$v^{\phi}$.  In terms of proper time, this is given by
\cite{LL75}
\begin{multline*}
  (v^{\phi})^2=e^{-\nu}r^2\left(\frac{d\Omega}{dt}\right)^2
  =e^{-\nu}r^2\left(\frac{d\Omega}{ds}\frac{ds}{dt}\right)^2\\
  =e^{-\nu}r^2\dot{\Omega}^2\frac{1}{\dot{t}}.
\end{multline*}
Substituting for $\dot{\Omega}^2$, we get from Eq. (\ref{E:L1})
\begin{equation}\label{E:tangential1}
  (v^{\phi})^2=\frac{e^{\nu}L^2}{r^2E^2}=\frac{1}{2}\,r\,\nu'.
\end{equation}
This expression can be integrated to yield
\begin{equation}\label{E:tangential2}
  e^{\nu}=B_0r^l,
\end{equation}
where $B_0$ is an integration constant and $l=2(v^{\phi})^2$.
Equivalently, the line element (\ref{line1}) can be written
\begin{equation}\label{E:line2}
  ds^2=-\left(\frac{r}{r_0}\right)^{2(v^{\phi})^2}dt^2
     +e^{\lambda(r)}dr^2+r^2d\Omega^2.
\end{equation}
As a result, the model has a well-defined Newtonian limit \cite
{BHL08}.  (Observe also that $V_{eff}(r)\sim 1/r^2$,
characteristic of dark matter, as a consequence of the flat
rotation curves.)

Now from Eq. (\ref{E:Veff})
\begin{equation}
  V_{eff}(r)_{rr}=\frac{6L^2}{r^4}-E^2e^{-\nu}(\nu')^2
     +E^2e^{-\nu}\nu''.
\end{equation}
By making the substitutions for $L^2$, $E^2$, and $\nu$, the
second derivative reduces to
\begin{equation}\label{E:Vfinal}
  V_{eff}(r)_{rr}=\frac{2l}{r^2}>0,
\end{equation}
confirming the existence of stable orbits (a detailed analysis of
which will be provided in the Sec. VI).

\section{Solutions}
In this section we determine $e^{\lambda(r)}$ in the line
element (\ref{line1}) by considering two cases, a
non-interacting two-fluid model and an interacting
two-fluid model.

\subsection {Non-interacting two-fluid model}
In our first case, the two fluids, normal matter and quark
matter, do not interact.  The resulting conservation equations
are therefore independent of each other.  Using
Eqs. (\ref{eq7}) and (\ref{eq8}), we have
\begin{equation}
 \frac{d\rho}{dr}   +
 \nu^\prime \left(\frac{1+m}{2m}\right) \rho = 0
    \label{conserve1}
\end{equation}
and
\begin{equation}
  \frac{d\rho_q}{dr}   +
 2\nu^\prime ( \rho_q -B) = 0  \label{conserve2}.
\end{equation}
The solutions are
\begin{equation}\label{E:sol1a}
\rho=\rho_{0}r^{-l(1+m)/2m}
\end{equation}
and
\begin{equation}\label{E:sol2a}
\rho_q=B+\rho_{0q} r^{-2l},
\end{equation}
where $\rho_0$ and $\rho_{0q}$ are integration
constants. Equations (\ref{eq4}), (\ref{eq5}), and
(\ref{eq6}) now yield
\begin{multline}\label{E:masterequation}
8\pi\left[ \rho(1+3m)+(2\rho_q-4B)\right]=\\
    e^{-\lambda}\left[-\frac{\lambda^\prime
 \nu^\prime}{2} + \frac{{(\nu^\prime})^2}{2} + \nu^{\prime \prime}
 +\frac{2\nu^\prime}{r} \right].
\end{multline}
This equation is linear in $e^{-\lambda}$ and leads to
\begin{equation}
  \frac{d}{dr}(e^{-\lambda}r^{2+l})=\\8\pi\left(\frac{2}{l}\right)
  r^{3+l}[\rho(1+3m)+2\rho_q-4B]
\end{equation}
by making use of Eq. (\ref{E:tangential2}). Incorporating Eqs.
(\ref{E:sol1a}) and (\ref{E:sol2a}), we now obtain
\begin{multline}\label{E:noninteracting-1}
e^{-\lambda} = D r^{-2-l} +\frac{2}{l} 8\pi \left[\frac{\rho_0
(1+3m) r^{2-l(1+m)/2m}}{4+l-l(1+m)/2m}\right.\\ +\left.
\frac{2\rho_{0q}r^{2-2l}}{4-l}-\frac{2B r^2}{4+l}\right],
\end{multline}
where $D$ is an integration constant.

\subsection {Interacting two-fluid model}
\noindent
In our second case, the two fluids are assumed to interact.  The
resulting conservation equations will then take the following
form:
\begin{equation}\label{E:conserve3}
  \frac{d\rho}{dr}   +
   \nu^\prime \left(\frac{1+m}{2m}\right) \rho = Q
\end{equation}
and
\begin{equation}\label{E:conserve4}
  \frac{d\rho_q}{dr}   +
  2\nu^\prime ( \rho_q -B) = -3Q.
\end{equation}
The quantity $Q$ expresses the interaction between
the dark-energy components. We assume that there is an energy
transfer from quark matter to normal matter. So a positive $Q$
is a natural choice that ensures that the second law of
thermodynamics is fulfilled \cite{PW09}. Moreover,
Eqs. (\ref{E:conserve3}) and (\ref{E:conserve4}) imply that
the interaction term is proportional to $r$ and vanishes as
$r\rightarrow \infty$. To meet these criteria, we assume that
\begin{equation}
Q \propto \frac{1}{r^n} \Rightarrow Q= \frac{Q_0}{r^n},
\end{equation}
where $0<n\le a$ for some $a<1$ and $Q_0$ is a constant of
proportionality. (This assumption is made strictly for
computational convenience, allowing us to write explicit
solutions; so no particular physical significance should
be attached to the range on $n$). The respective solutions
of Eqs. (\ref{E:conserve3}) and (\ref{E:conserve4}) are now
given by
\begin{equation}\label{E:sol1b}
  \rho=\rho_{0(in)}r^{-l(1+m)/2m}
  +\frac{Q_0r^{1-n}}{1-n+l(1+m)/2m}
\end{equation}
and
\begin{equation}\label{E:sol2b}
 \rho_q=B  +\rho_{0q
 (in)}r^{-2l} - \frac{3Q_0r^{1-n}}{1-n+2l},
\end{equation}
where $\rho_{0 (in)}$ and $\rho_{0q (in)}$
 are integration constants.

The solution of Eq. (\ref{E:masterequation}) now becomes
\begin{multline}\label{E:interacting-1}
e^{-\lambda} = Fr^{-2-l} +\frac{2}{l} 8\pi\left[
\frac{\rho_{0(in)}(1+3m)r^{2-l(1+m)/2m}}{4+l-l(1+m)/2m}\right.\\
\left.+\frac{2\rho_{0q(in)}r^{2-2l}}{4-l}-\frac{2Br^2}{4+l}\right]\\
+\frac{2}{l} 8\pi \left[\frac{Q_0
(1+3m)r^{3-n}}{[1-n+l(1+m)/2m](5-n+l)}
   \right.\\
-\left.\frac{6Q_0 r^{3-n}}{(1-n+2l)(5-n+l)}\right],
\end{multline}
where $F$ is an integration constant. Observe that apart from the
integration constants, the first part of solution
(\ref{E:interacting-1}) is the same as solution
(\ref{E:noninteracting-1}).

\section{On the question of an attractive gravity}
We may consider the question of an attractive gravity in the halo
region by studying the geodesic equation
\begin{equation}\label{E:geodesic}
           \frac{d^2x^\alpha}{d\tau^2}
  +\Gamma_{\mu\gamma}^{\alpha}\frac{dx^\mu}
   {d\tau}\frac{dx^\gamma}{d\tau}= 0.
\end{equation}
This equation implies that \cite{Rahaman2012}
\begin{equation}
  \frac{d^2 r} {d\tau^2} = - \frac{1}{2} e^{-\lambda}
    \left[\frac{d }{dr}e^\lambda \left(\frac{dr}{d\tau}
\right)^2 + \frac{d }{dr}e^\nu\left(\frac{dt}{d\tau}
\right)^2\right].
\end{equation}
Since $\dot{r}=0$, as before, we get from Eq. (\ref
{E:tangential2}),
\begin{equation}\label{E:attractive}
  \frac{d^2 r} {d\tau^2} = - \frac{1}{2} e^{-\lambda}
   B_0lr^{l-1}\left(\frac{dt}{d\tau}\right)^2<0
\end{equation}
as long as $e^{-\lambda}>0$, thereby yielding an attractive gravity
in the halo region.

In the non-interacting case related to Eq. (\ref{E:noninteracting-1}), the
dominant terms are the last two because of the size of the bag
constant $B$. We can see from the EoS-1 in Eq. (\ref{eq7}) that
\begin{equation}\label{E:EoS2}
     p_q=\frac{1}{3}(\rho_{0q}r^{-2l}-3B).
\end{equation}
So whenever the pressure $p_q$ is positive, $\rho_{0q}
>3Br^{2l}>3B$.  It follows that
\begin{equation}\label{E:gravityeff}
  e_{eff}^{-\lambda}=\frac{2\rho_{0q}r^{2-2l}}{4-l}
    -\frac{2Br^2}{4+l}
\end{equation}
is positive and since $\rho_0\ll\rho_q$ and
$l\approx 0.000001$ \cite{Nandi2009b}, $e^{-\lambda(r)}$
is also positive.

The same argument can be applied to the interacting case
because of the similarity between
Eqs. (\ref{E:noninteracting-1}) and ({\ref{E:interacting-1}).
So we obtain an attractive gravity in both cases.

The conclusion is somewhat different if $p_q$ is negative. Based on
Eq. (\ref{E:EoS2}), $\rho_{0q}$ can still be large enough to yield
$e_{eff}^{-\lambda}>0$. Now suppose that the corresponding expression
in Eq. (\ref{E:interacting-1}) is also positive. At a first glance
it seems difficult to quantify precisely the structure of quark matter
as much of it even now remains conjectural. However, for
certain combinations of values of the
various parameters, the extra terms involving $Q_0$ will produce
a negative value for the total. So we no longer have an attractive
gravity for the interacting case. Of course, if the pressure $p_q$
is negative and sufficiently large in absolute value, then
$e^{-\lambda(r)}<0$ in both cases, so that quark matter
would not be a suitable model.

\section{Analysis of the Stability of circular orbits}
In the Introduction we have assumed the circular orbits as an input
of flat rotation curve and have shown stable via the Eq. (\ref{E:Vfinal}).
Here we analyze the stability of the orbits
in a more detailed form to distinguish between the interacting
and non-interacting cases. Let a test particle with four velocity
$U^{\alpha}=\frac{dx^{\sigma}}{d\tau}$ moving in the region of
spacetime given in (23). Assuming $\theta=\pi/2$, the equation
$g_{\nu\sigma}U^{\nu}U^{\sigma}=-m_{0}^{2}$ yields
\begin{equation}
\left( \frac{dr}{d\tau}\right)  ^{2}=E^{2}+V(r),
\end{equation}
with
\begin{equation}
V(r)=-\left[E^{2}\left( 1-\frac{r^{-l}e^{-\lambda} }{B_{0}}\right)
+e^{-\lambda}\left( 1+\frac{L^{2}}{r^{2}}\right) \right].
 \end{equation}

Here the two conserved quantities, namely relativistic energy (E)
and angular momentum (L) per unit rest mass of the test particle
respectively are
\begin{equation}
E=\frac{U_{0}}{m_{0}}\quad and \quad L=\frac{U_{3}}{m_{0}}.
\end{equation}

If the circular orbits are defined by $r=R$, then
$\frac{dR}{d\tau}=0$ and, additionally,
$\frac{dV}{dr}\mid_{r=R}=0$. Above two conditions result
\begin{equation}
L=\pm\sqrt{\frac{l}{2-l}}R
\end{equation}
and, using $L$ in $V(R)=-E^{2}$, we get
\begin{equation}
E=\pm\sqrt{\frac{2B_{0}}{2-l}}R^{l/2}.
\end{equation}
The orbits will be stable if $\frac{d^{2}V}{dr^{2}}\mid_{r=R}<0$
and unstable if $\frac{d^{2}V}{dr^{2}}\mid_{r=R}>0$.

By putting the expressions for $L$ and $E$ in
$\frac{d^{2}V}{dr^{2}}\mid_{r=R}$ and then by using the
Eq. (\ref{E:interacting-1}) of the interacting case, finally we get
\begin{multline}\label{E:interacting-2}
\frac{d^{2}V}{dr^{2}}\mid_{r=R}=-  \left[ \frac{lR^2}{2-l} \{
F(4+l)(5+l)R^{-6-l}  \right.\\ \left.+F_1 \frac{l(1+m)}{2m} ( 1+
\frac{l(1+m)}{2m} ) R^{-2-\frac{l(1+m)}{2m}}
 \right.\\
\left.+  F_2 2l(1+2l)R^{-2-2l} + F_4 n(1-n) R^{-1-n} \} \right.\\
\left.+ F(2+l)(3+l)R^{-4-l} \right.\\ \left.+ F_1
(2-\frac{l(1+m)}{2m })(1-\frac{l(1+m)}{2m })
R^{-\frac{l(1+m)}{2m}} \right.\\ \left. + F_2(2-2l)(1-2l) R^{-2l}
+ 2F_3 +F_4 (3-n)(2-n) R^{1-n} \right.\\ \left.  +
\frac{2R^l}{2-l} \{ F(2+2l)(3+2l) R^{-4-2l} + \right.\\ \left. F_1
(2-\frac{l(1+m)}{2m })(1-\frac{l(1+m)}{2m })
R^{-l-\frac{l(1+m)}{2m}} \right.\\ \left.+F_2(2-2l)(1-2l) R^{-3l}
+\right.\\ \left.F_3 (2-l)(1-l)R^{-l} +F_4 (3-n-l)(2-l-n)
R^{1-n-l} \}\right],\\
 \end{multline}
 where\\
 $F_1 =  8\pi \left(\frac{2}{l}\right) \left[ \frac{\rho_{0(in)}(1+3m)}{ 4+l-\frac{l(1+m)}{2m}}  \right]
$, \\
 $F_2 =  8\pi \left(\frac{2}{l}\right) \left[\frac{2\rho_{0q(in)}}{4-l}\right] $,\\
$ F_3 =  8\pi \left(\frac{2}{l}\right) \left[\frac{2B}{4+l}\right] $,\\
$ F_4 =  8\pi \left(\frac{2}{l}\right) \left[\frac{Q_0
(1+3m)}{[1-n+l(1+m)/2m](5-n+l)} -\frac{6Q_0
}{(1-n+2l)(5-n+l)}\right]. $
We note that for non-interacting case when $Q_0 =0$,  $\frac{d^{2}V}{dr^{2}%
}\mid_{r=R}<0$ since, $l\approx 10^{-6}$ and $m , n < 1$. Thus the
circular orbits are always stable for non-interacting situation.
When, $Q_0 \neq 0$ i.e. for interacting two fluids, $F_4 < 0$.
However, if $Q_0$ is very small then,  $\frac{d^{2}V}{dr^{2}%
}\mid_{r=R}$ is also negative and circular orbits are stable.
The last condition implies that the interaction is very weak.
This weak interaction is a characteristics of quark matter at
higher temperatures. Thus, in the Weakly Interacting Massive Particle
(WIMP) category of the dark-matter halo, besides neutrino-like light
particles, quark matter may be one of the possible candidates, which,
in turn, has important implications for the galactic structure
formation and evolution.

\section{Behavior of quark matter: a dual role}
The above results may have an interesting interpretation on a
cosmological scale. If $\rho_q$ and $p_q$ refer to the matter
content in the galactic halo region, then, as we have seen,
if $p_q>0$, then bag-model quark matter behaves like dark
matter. Now suppose that $\rho_q$ refers to the energy
density obtained by including a region beyond the halo
large enough so that  the resulting reduced value leads
to $p_q<0$ in Eq. (\ref{E:EoS2}).  Assume also that
$l=2(v^{\phi})^2$ decreases so gradually in the outward
radial direction that we may treat it as a constant.
Substituting the expression for $p_q$ from Eq.
(\ref{E:EoS2}) in the Friedmann equation
$(d^2a(t)/dt^2)/a(t)=-\frac{4\pi}{3}(\rho+3p)$,
we obtain
\begin{equation}
  \frac{1}{a(t)}\frac{d^2a(t)}{dt^2}
  =-\frac{4\pi}{3}(\rho_q+\rho_{0q}r^{-2l}-3B).
\end{equation} Sufficiently small $\rho_q$ and $\rho_{0q}$ now
result in a positive acceleration, in the manner of dark energy.
It seems therefore on the galactic level, quark matter behaves
like dark matter and on the global level like dark energy.

\section{Conclusion}
By taking the flat rotation curves as input and treating the matter
content in the galactic halo region as quark matter, we obtained
a spacetime metric for the galactic halo.

One can see that the solutions given in the Eq.
(\ref{E:noninteracting-1}) and Eq. (\ref{E:interacting-1}) are the
interior solution of the spacetime metric given by the Eq.
(\ref{E:line2}) which is neither asymptotically flat nor a
spacetime due to a centrally symmetric black hole. The metric
describes that region for which the tangential velocity of the
test particle is constant.

Since flat rotation curves were assumed, the problem of
determining the suitability of quark matter as a model for dark
matter was reduced to finding conditions under which gravity in
the halo region is attractive.

We used the MIT bag model whose equation of state is
$p_q=\frac{1}{3}(\rho_q-4B)$ and considered two cases, non-interacting
and interacting two-fluid models (referring to quark matter and
baryonic matter). In both cases, if the pressure $p_q>0$,
gravity in the halo is attractive. If $p_q<0$, with a sufficiently large
absolute value, then gravity in the halo is repulsive. There exist
values of $p_q$ between these extremes in which gravity is
attractive in the non-interacting case but not in the interacting
case.

We have also investigated the stability of circular orbits and
shown that the circular orbits are always stable for
non-interacting situation. However, interacting two fluids
circular orbits are found to be stable under certain conditions
which imply that the interaction is very weak.

In our investigation it is also revealed that on the galactic level,
quark matter behaves like dark matter whereas on the global level like
dark energy.

In summary, however, quark matter fitting the MIT bag model is indeed a
suitable model for dark matter under fairly general conditions.

As an additional remark, in the interacting case between quark
and neutral hydrogen a rise of temperature is expected to happen.
If this environment of increased temperature in the interstellar
medium is detectable observationally then the present model is
physically viable one. The aspect of thermodynamical processes
involved in this phenomenon can be undertaken in a future project.

\section*{Acknowledgments}
FR and SR are thankful to the authority of The Inter-University
Centre for Astronomy and Astrophysics, Pune, India for providing
Visiting Associateship programme under which a part of this work
was carried out. We are grateful to an anonymous referee for
suggesting some pertinent issues that have led to significant
improvements. Special thanks are due to Dr. A. K. Jafry for
helpful discussions on the referee's comments. FR personally
is also thankful to PURSE, DST and UGC, Government of India
for providing financial support under Research Award scheme.


\begin{thebibliography}{99}

\bibitem{jO32} J. Oort, Bull. Astron. Inst. Neth. VI \textbf{249} (1932) 249.
\bibitem{fZ33} F. Zwicky, Helvet. Phys. Acta \textbf{6} (1933) 110.
\bibitem{fZ37} F. Zwicky, Astrophys. J. \textbf{86} (1937) 217.
\bibitem{Rahaman2008} F. Rahaman, et al., Mon. Not. R. Astron. Soc. \textbf{389} (2008) 27.
\bibitem{Nandi2009a} K.K. Nandi, et al., Mon. Not. R. Astron. Soc. \textbf{399} (2009) 2079.
\bibitem{Rahaman2012} F. Rahaman et al., Gen. Relativ. Gravit. \textbf{44} (2012) 905.
\bibitem{Matos2000} T. Matos, F.S. Guzm{\'a}n and D. Nu{\~n}ez, Phys. Rev. D \textbf{62} (2000) 061301.
\bibitem{Nandi2009b} K.K. Nandi, I. Valitov, and N.G. Migranov, Phys. Rev. D \textbf{80} (2009) 047301.
\bibitem{Rahaman2010} F. Rahaman, et al., Phys. Lett. B \textbf{694} (2010) 1015l2.
\bibitem{ESM90} G. Efstathiou, W. Sutherland, and S.J. Madox, Nat. \textbf{348} (1990) 705.
\bibitem{aP04} A.C. Pope, et al., Astrophys. J. \textbf{607} (2004) 655.
\bibitem{Khlopov2011} M. Yu. Khlopov, Mod. Phys. Lett. A, \textbf{26} (2011) 2823.
\bibitem{Heinz2000} U. Heinz and M. Jacob, arXiv:nucl-th/0002042.
\bibitem{Schaefer2003} T. Schaefer, in The Proceedings of the BARC Workshop on Quarks and Mesons, arXiv:hep-ph/0304281 (2003).
\bibitem{Muller2005} B. M{\"u}ller, arXiv:nucl-th/0508062.
\bibitem{PSS} M.A. Perez-Garcia, J. Silk and J.R. Stone, Phys. Rev. Lett. \textbf{105} (2010) 141101.
\bibitem{Forbes2010} M.M. Forbes, K. Lawson and A.R. Zhitnitsky, Phys. Rev. D  \textbf{82} (2010) 083510.
\bibitem{Weissenborn2011} S. Weissenborn et al., Astrophys. J. \textbf{740} (2011) L14.
\bibitem{Drake2002} J.J. Drake, Astrophys. J. \textbf{572} (2002) 996.
\bibitem{Jaikumar2006} P. Jaikumar, S. Reddy and A.W. Steiner, Phys. Rev. Lett. \textbf{96} (2006) 041101.
\bibitem{Rodrigues2011} H. Rodrigues, S. Barbosa Duarte and J.C.T. de Oliveira, Astrophys. J. \textbf{730} (2011) 31.
\bibitem{Bordbar2011} G.H. Bordbar and A.R. Peivand, arxiv:1103.1250.
\bibitem{Madsen1999} J. Madsen, Lect. Notes Phys. \textbf{516} (1999) 162.
\bibitem{Weber2005} F. Weber, Prog. Part. Nucl. Phys. \textbf{54} (2005) 193.
\bibitem{Bhattacharyya1999} A. Bhattacharyya, et al., Nucl. Phys. A {\bf 661} (1999) 629.
\bibitem{Bhattacharyya2000} A. Bhattacharyya, et al., Phys. Rev D {\bf 61} (2000) 083509.
\bibitem{Bhattacharyya2003} A. Bhattacharyya, et al., Pramana {\bf 60} (2003) 909.
\bibitem{Witten1984} E. Witten, Phys. Rev D {\bf 30} (1984) 272.
\bibitem{EMU1991} EMU01 Collab., M.I. Adamovich, et al., Phys. Lett. B {\bf 263} (1991) 539.
\bibitem{Fromerth2005} M. Fromerth and J. Rafelski, {\it CINPP Conference}, Kolkata,
February 2005, page 22.
\bibitem{Madsen1986} J. Madsen, H. Heiselberg and K. Riisager, Phys. Rev. D {\bf 34} (1986) 2974.
\bibitem{Madsen1991} J. Madsen and M. L. Olesen, Phys. Rev. D {\bf 43} (1991)
1069.
\bibitem{Bhattacharjee1993} P. Bhattacharjee, et al., Phys. Rev. D {\bf 48} (1993) 4630.
\bibitem{Sumiyoshi1991} K. Sumiyoshi and T. Kajino, Nucl. Phys. {\bf 24} (1991) 80.
\bibitem{Walker1991} T. Walker, et al., Astrophys. J. {\bf 376} (1991)
51.
\bibitem{Chodos1974} A. Chodos, et al., Phys. Rev. D \textbf{9} (1974) 3471.
\bibitem{Rahaman2011} F. Rahaman, et al., Eur. Phys. J. C, DOI: 10.1140/epjc/s10052-012-2071-5, arXiv:1108.6125.
\bibitem{BHL08} C.G. B{\"o}hmer, T. Harko and F. S. N. Lobo,
   Astropart. Phys. \textbf{29} (2008) 386.
\bibitem{NSS01} U. Nucamendi, M. Salgado and D. Sudarsky, Phys. Rev. D
   \textbf{63} (2001) 125016.
\bibitem{LL75} L.D. Landau and E.M. Lifshitz, \emph{The Classical Theory of Fields},
   Oxford, Pergamon Press (1975).
\bibitem{PW09} D. Pavon and B. Wang, Gen. Relativ. Gravit. \textbf{41} (2009) 1.

\end{thebibliography}
\end{document}